\documentclass[twocolumn,amsmath,amssymb,aps,pra,floatfix,nofootinbib,superscriptaddress]{eptcs}
\usepackage{graphicx,graphics,times,color}
\usepackage{bm}
\usepackage{longtable}

\def\be{\begin{equation}}
\def\ee{\end{equation}}
\def\ba{\begin{eqnarray}}
\def\ea{\end{eqnarray}}
\def\la{\langle}
\def\ra{\rangle}

\title{Engineering Long Range Distance Independent Entanglement through Kondo Impurities in Spin Chains}
\author{Abolfazl Bayat 
\institute{Department of Physics and Astronomy,\\ University
College London, Gower St.,\\ London WC1E 6BT, UK}
\and 
Pasquale Sodano
\institute{Dipartimento di Fisica e Sezione I.N.F.N.,\\
Universita' di Perugia, Via A. Pascoli,\\ Perugia, 06123, Italy}
\and
Sougato Bose
\institute{Department of Physics and Astronomy,\\ University
College London, Gower St.,\\ London WC1E 6BT, UK}
}
\def\titlerunning{Engineering Long Range Distance Independent Entanglement}
\def\authorrunning{A. Bayat, P. Sodano  \& S. Bose}

\begin{document}

\maketitle

\begin{abstract}
We investigate the entanglement properties of the Kondo spin chain
when it is prepared in its ground state as well as its dynamics
following a single bond quench. We show that a true measure of
entanglement such as negativity enables to characterize the unique
features of the gapless Kondo regime. We determine the spatial
extent of the Kondo screening cloud and propose an ansatz for the
ground state in the Kondo regime accessible to this spin chain; we
also demonstrate that the impurity spin is indeed maximally
entangled with the Kondo cloud. We exploit these features of the
entanglement in the gapless Kondo regime to show that a single
local quench at one end of a Kondo spin chain may always induce a
fast and long lived oscillatory dynamics, which establishes a high
quality entanglement between the individual spins at the opposite
ends of the chain. This entanglement is a footprint of the
presence of the Kondo cloud and may be engineered so as to attain
- even for very large chains- a constant high value independent of
the length; in addition, it is thermally robust. To better
evidence the remarkable peculiarities of the Kondo regime, we
carry a parallel analysis of the entanglement properties of the
Kondo spin chain model in the gapped dimerised regime where these
remarkable features are absent.
\end{abstract}


\section{Introduction}\label{introduction}

Kondo systems \cite{kondo,Sorensen-Affleck} are expected to be
very distinctive in the context of entanglement for at least two
reasons:  (a) Despite being ``gapless", they support the emergence
of a length scale $\xi$ \cite{kondo,Sorensen-Affleck} which should
be reflected in the entanglement, making it markedly different
from that in the more conventional gapless models studied so far;
(b) They are expected to have a more exotic {\em form} of
entanglement than the widely studied spin-spin and complementary
block entanglements. Indeed, in Kondo systems, the impurity spin
is expected to be mostly entangled with only a specific block of
the whole system. This is, of course, merely an intuition which
needs to be quantitatively verified with a genuine measure of
entanglement: this task has been accomplished in \cite{Kondous}
where we provided the only characterization of the Kondo regime
based entirely on a true measure of entanglement such as
negativity \cite{negativity}.

The simplest Kondo model \cite{kondo, hewson} describes a single
impurity spin interacting with the conduction electrons in a
metal; the ground state is a highly nontrivial many body state in
which the impurity spin is screened by conduction electrons in a
large orbital of size $\xi$- the so called Kondo screening length.
Many physical observables vary on the characteristic length scale
$\xi$, which is a well defined function of the Kondo coupling
\cite{kondo}. The screening length $\xi$ determines the spatial
extent of the Kondo cloud whose signatures in physical systems
have been so far a challenging problem repeatedly addressed by
various means \cite{Sorensen-Affleck,Frahm,dots}. As we shall
detail in the following an entanglement measure enables to fully
determine $\xi$ \cite{Kondous} and to use its knowledge to
engineer long range distance independent entanglement
\cite{Sodano}.

Recently \cite{kondo-Affleck}, it has been pointed out that the
universal low energy long distance behavior of this simple Kondo
model arises also in a spin chain when a magnetic impurity is
coupled to the end of a gapless Heisenberg anti-ferromagnetic
$J_1-J_2$ spin 1/2 chain. The spin chain Kondo model
\cite{kondo-Affleck} is defined by the Hamiltonian
\begin{eqnarray}\label{Hamiltonian}
    H_I&=&J'(J_1\sigma_1.\sigma_2+J_2\sigma_1.\sigma_3)\cr
    &+&\sum_{i=2}^{N-1}J_1\sigma_i.\sigma_{i+1}+J_2\sum_{i=2}^{N-2}\sigma_i.\sigma_{i+2},
\end{eqnarray}
where $\sigma_i=(\sigma_i^x,\sigma_i^y,\sigma_i^z)$ is a vector of
Pauli operators at site $i$, $N$ is the total length of the chain,
$J_2$ is the next nearest neighbor coupling and the nearest
neighbor coupling $J_1$ is normalized to $1$. The impurity spin,
located at one end of the chain, is accounted for by weaker
couplings to the rest of the system; in the following, see Fig.
\ref{fig1}a, both couplings $J_1$ and $J_2$ are weakened by the
same factor $J'$, which quantifies then the impurity strength. for
$0\leq J_2 \leq J_2^c=0.2412$, the spin system is gapless and it
supports a Kondo regime
\cite{Laflorencie-Affleck,Sorensen-Affleck}. For $J_2>J_2^c$, the
system enters the gapped \emph{dimer regime}, where the ground
state takes a dimerised form; at the Majumdar-Ghosh \cite{MG}
point ($J_2=0.5$), the ground state becomes just a tensor product
of singlets. For $J_2>0.5$, incommensurability effects
\cite{pertinentAffleck} emerge.

In the following we detail our analysis \cite{Kondous,Sodano} of
the remarkable features of entanglement in the Kondo regime
\cite{kondo,Sorensen-Affleck} of the spin chain Kondo model. First
of all, we use negativity \cite{negativity} to characterize
\cite{Kondous} the entanglement in the Kondo regime; namely,for
this spin chain in the Kondo regime: (i) we demonstrate that the
impurity spin is maximally entangled with the Kondo cloud; (ii) we
determine the spatial extent of the Kondo screening length $\xi$
using only an entanglement measure; (iii) we motivate an ansatz
for the ground state in the Kondo regime; (iv) we evidence the
scaling of negativity as pertinent parameters are varied. Then, we
use the knowledge acquired in \cite{Kondous} to engineer- for
chains of arbitrary size $N$- a long range distance-independent
entanglement through a non-perturbative dynamical mechanism which
requires only a \textit{minimal action} on a spin chain, namely a
sudden quench of a single bond \cite{Sodano}. The ensuing long
range distance independent entanglement- as well as the remarkable
entanglement oscillations observed in our numerical simulations
\cite{Sodano}- are a footprint of the emergence of the length
scale $\xi$ in an -otherwise gapless- system. To accomplish these
tasks we designed \cite{Kondous} a Density Matrix Renormalization
Group (DMRG) approach enabling to investigate the entanglement
between a single spin and a pertinent block of the chain. We also
developed \cite{Sodano} a time dependent DMRG to simulate the
dynamics of the system after a sudden quench of a local bond; to
better evidence the unique properties of the entanglement in the
Kondo regime we carried a parallel analysis of the entanglement
properties of this model in the gapped dimerised regime where all
these remarkable features of entanglement are absent.

 The investigation of entanglement in many-body condensed matter systems is currently a topic of intense
 activity \cite{Amico-RMP,bose-vedral,fazio-osborne,vidal,korepin,Calabrese,Laflorencie-Affleck,Sorensen-Affleck,Lehur,Kondous,Sodano}.
In many instances investigations have focussed on the entanglement
between individual elements, such as single spins, or the
entanglement between two complementary blocks in the ground state
of a condensed matter system. The former is generically non-zero
only between nearest or next to nearest neighbors
\cite{bose-vedral,fazio-osborne}; at variance, for complementary
blocks, the whole system is in a {\em pure state} and the von
Neumann entropy is a {\em permissible} measure of the
entanglement. In conventional gapless phases, due to the absence
of an intrinsic length scale, the von Neumann entropy diverges
with the size of the blocks \cite{vidal,korepin,Calabrese}; at
variance, our analysis \cite{Kondous,Sodano} provides for the
first time a characterization of the entanglement in gapless
regimes of a many body system where a length scale emerges as a
result of the presence of a Kondo impurity.

From a more practical viewpoint there is now a huge demand for
long range entanglement between individual spins in quantum
information theory. For instance, the long range entanglement
between individual particles gives the opportunity of implementing
teleportation \cite{teleportation} which offers perfect quantum
communication between distant parties. Though entanglement in
condensed matter systems is typically very short ranged
\cite{fazio-osborne}, there have been a few other proposals for
long distance entanglement; however, they come at a high price.
For instance, there are proposals exploiting weak couplings of two
distant spins to a spin chain
\cite{lorenzo1-illuminati,lorenzo2-plenio-li-wojcik}, which have
very limited thermal stability or very long time-scale of
entanglement generation. Otherwise, a dynamics has to be induced
by large-scale changes to the Hamiltonian of a system
\cite{Hannu-Hangii}.

The paper is organized as follows: In section (\ref{static}) we
review our analysis \cite{Kondous} of the static entanglement
properties of the ground state of the Kondo spin chain model using
negativity\cite{negativity} as a true measure of entanglement:
there, using only quantum information tools, we are
able\cite{Kondous} to determine the spatial extent of the Kondo
cloud as well as to provide an ansatz for the ground state in the
Kondo regime of the spin chain Kondo model. In section
(\ref{dynamics}) we point out \cite{Sodano} how - only in the
Kondo regime of the spin chain Kondo model- a local quench of the
last bond of the Kondo spin chain induces a thermally robust high
quality long range and distance-independent entanglement between
the two individual ending spins of the chain: there, we emphasize
how the onset of this long range distance-independent entanglement
may be regarded as the footprint of the emergence of the Kondo
cloud in the Kondo regime of the spin chain Kondo model. Section
(\ref{summary}) is devoted to a short discussion of our results.

\section{Entanglement in the Ground State of the Spin Chain Kondo Model} \label{static}

In this section we detail the results of our analysis
\cite{Kondous} of the entanglement properties of the ground state
of the Hamiltonian described by (\ref{Hamiltonian}). We show that
a true measure of entanglement enables one to use only quantum
information tools to determine the Kondo screening length $\xi$
and to provide an ansatz for the ground state of this spin chain
in the Kondo regime.

A true measure of entanglement should satisfy a set of postulates
- for example, it should be non-increasing under local actions:
such a genuine measure does exist for two sub-systems of arbitrary
size even when their combined state is mixed, as it happens in
Kondo systems. This measure is the {\em negativity}
\cite{negativity} and it has been successfully used to quantify
the entanglement in a harmonic chain \cite{audenaert,kim-vedral}
and between distant regions of critical systems
\cite{hannu,alex-retzker}. For bipartite systems, negativity is

defined as $E=\sum_i|a_i|-1$, where $a_i$ denote the eigenvalues
of the partial transpose of the whole density matrix of the system
with respect to one of the two subsets of the given partition and
$|...|$ is the absolute value \cite{negativity}.

In the following we use DMRG as the numerical tool needed to
measure the static entanglement properties observed in the Kondo
regime of the spin chain Kondo model.

\subsection{The Entanglement Healing Length}

\begin{figure}
\centering
    \includegraphics[width=7cm,height=6cm,angle=0]{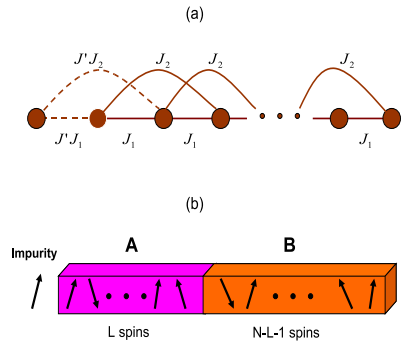}
    \caption{ (a) Kondo Spin chain with next nearest neighbor Heisenberg interaction with one impurity at one end.
    (b) The chain is divided into three parts, an impurity, a block $A$ and a block $B$. Entanglement is computed
    between the impurity and block $B$.}
     \label{fig1}
\end{figure}

To study the entanglement of the ground state we divide- see Fig.
\ref{fig1}b - all the spins of the chain in three different
groups: the impurity spin, block $A$, which contains the $L$ spins
next to the impurity ($L=0,1,...,N-1$) and block $B$ formed by the
remaining $N-L-1$ spins. We used \cite{Kondous} negativity to
fully characterize the entanglement between the impurity and block
$B$ in both the gapless Kondo and the gapped dimerised regimes.

We determined \cite{Kondous} the size of the block $A$ when the
entanglement between the impurity and block $B$ is almost zero; by
this procedure one defines \cite{Kondous} an Entanglement Healing
Length (EHL) $L^*$, i.e. the length of the block $A$ which is
maximally entangled with the impurity. It should be emphasized
that an EHL may be defined for all spin chains: however, we showed
\cite{Kondous} that, only in the gapless Kondo regime of the Kondo
spin chain, EHL scales with the strength of the impurity coupling
just as the Kondo screening length, $\xi$, does.

In the gapless regime of the Kondo spin chain, measuring the EHL
through negativity \cite{Kondous} yields a genuine quantum
information tool to detect the Kondo screening length
\cite{Sorensen-Affleck,Frahm,dots}. In addition we {\em found}
that- in the Kondo regime- entanglement, as quantified by
negativity, is a homogeneous function of the two ratios: $N/L^*$
and $L/N$, where $L$ is the size of the block $A$, i.e. the block
adjacent to the impurity, and $N$ is the length of the whole
chain. As a result, the entanglement in the Kondo regime is
essentially unchanged if one rescales all the length scales with
the EHL $L^*$. Of course, EHL can be defined also in the gapped
dimerised regime but negativity is now a function of the three
independent quantities $N$, $L$ and $L^*$: i. e., scaling of
entanglement with EHL is absent in the gapped dimerised regime
\cite{Kondous}.

\begin{figure}
\centering
    \includegraphics[width=7cm,height=6cm,angle=0]{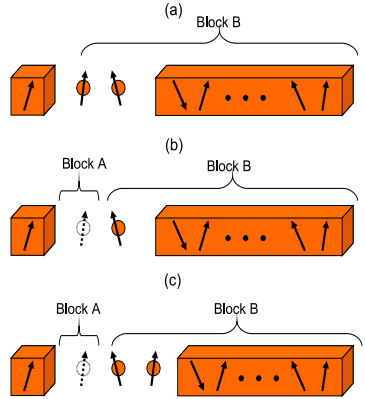}
    \caption{(a) DMRG representation of the state of the chain keeps two intermediate spins
    in ordinary computational basis and the left and right
    blocks in a truncated DMRG basis. (b) The intermediate spin next to the impurity is traced out from the density matrix of the
    chain. This tracing is equivalent to adding the traced out spin
    to block $A$. (c) The basis of the right block of DMRG representation is
transformed so that a single spin in the left side of the
    right block is represented in the computational basis while the state of the new
right block is given in a DMRG basis.}
     \label{fig2}
\end{figure}

\subsection{Numerical Approach: Density Matrix Renormalization Group } \label{methodology}
  The DMRG \cite{white-DMRG} approach has been used \cite{Kondous} to compute the ground state of
the spin chain Kondo model for large chains up to $N=250$; in
order to avoid finite size effects, we took $N$ to be even and,
thus, avoided problems arising from accidental degeneracies.

In a DMRG approach the ground state of the system is partitioned
in terms of states of a left block, a right block (not to be
confused with blocks $A$ and $B$) and two intermediate spins as
shown in Fig. \ref{fig2}a. The states of the intermediate spins
are given in the computational ($|\uparrow\ra,|\downarrow\ra$)
basis, while the states of the both blocks are usually in some
non-trivial truncated DMRG basis. In this approach one has several
representations for the ground state which vary due to the number
of spins in the left (right) block and it is possible to go from
one representation to the other by applying pertinent operators on
each block.

The main issue in the DMRG is that the dimension of the left
(right) block is kept constant and independent of whatever spins
are there in that block. To have a fixed dimension for the left
(right) block we truncated the Hilbert space so that the amount of
entanglement between the two parts of the chain remains almost
unchanged \cite{white-DMRG}. To improve the precision of our
results we swept \cite{Kondous} all representations of the ground
state for several times to get the proper basis for the left and
the right blocks of all representations. After some sweeps, when
the ground state energy converges (we kept states for which the
error on the energy is less than $10^{-6}$), we paused to compute
the entanglement. We took a representation of the ground state in
which the left block contains just the single impurity spin and
the right block contains $N-3$ spins: as a result, the single
impurity spin is given in the computational basis and this allowed
us to compute the negativity later.

 From the DMRG state, one should trace
out the spins belonging to block $A$
 before computing the entanglement between the
impurity and block $B$ since it is most convenient to compute the
entanglement between the impurity and the block $B$: due to the
entanglement monogamy, this provides an equivalent information
about the entanglement of the impurity with the block $A$. Our
tracing procedure started with the density matrix of the ground
state of the system in the representation shown in Fig.
\ref{fig2}a; at this stage, the number of spins in the block $A$
is zero (no spin has been traced out), all spins except the
impurity belong to the block $B$, and the entanglement between the
impurity and the block $B$ is maximal (i.e. $E=1$). Then, we
traced out the intermediate spin next to the impurity as shown in
Fig. \ref{fig2}b; this amounts to putting that spin into block
$A$. Finally, as shown in Fig. \ref{fig2}c, we transformed the
DMRG basis of the right block so as to put a single spin at the
left of the right block in the computational basis, while the
state of the new right block is given in a DMRG basis. As a
consequence, the resulting density matrix  had the exact form of
Fig. \ref{fig2}a and we continued the procedure to trace one spin
at each step (i.e., put more spins in the block $A$) and computed
the entanglement between the impurity and block $B$.

\begin{figure}
\centering
    \includegraphics[width=8cm,height=7cm,angle=0]{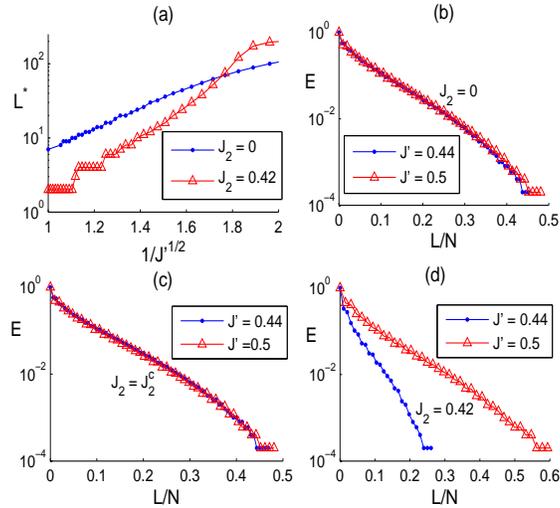}
    \caption{(a) $L^*$ vs. $1/\sqrt{J'}$ for both Kondo ($J_2=0$) and dimer regime ($J_2=0.42$).
    (b) Entanglement vs. $L/N$ for fixed $N/L^*=4$ when $J_2=0$.
    (c) Entanglement vs. $L/N$ for fixed $N/L^*=4$ at the critical point $J_2=J_2^c$.
    (d) Entanglement vs. $L/N$ for fixed $N/L^*=4$ in the dimer regime ($J_2=0.42$).}
     \label{fig3}
\end{figure}

\subsection{Scaling of Negativity and Ansatz for the Ground State in the Kondo Regime} \label{results}
As evidenced in \cite{Kondous}, there is an EHL $L^*$ so that, for
$L>L^*$, the entanglement between the impurity and block $B$ is
almost zero: $L^*$ provides us with an estimate of the distance
for which the impurity is mostly entangled with the spins
contained in block $A$. For large chains ($N>200$) in the Kondo
regime, one finds that $L^*$ is almost independent of $N$ and
depends only on $J'$. In the Kondo regime, i.e. for $J_2<J_2^c$,
$L^*$ depends on $J'$ just as the Kondo screening length $\xi$
does \cite{Laflorencie-Affleck,Sorensen-Affleck}; for small $J'$,
$L^*\propto e^{\alpha/\sqrt{J'}}$, where $\alpha$ is a constant.
We plot $L^*$ as a function of $1/\sqrt{J'}$ in Fig. \ref{fig3}a.
In a semi-logarithmic scale, the straight line plot exhibited in
the Kondo regime ($J_2=0$) shows that $L^*$ may be indeed regarded
as the Kondo screening length. Moreover, the nonlinearity of the
same plot in the dimer regime ($J_2=0.42$), especially for small
$J'$, shows that, in the gapped dimerised regime accessible to the
Kondo spin chain model, no exponential dependence on $1/\sqrt{J'}$
holds.

 There is \cite{Kondous} also a remarkable scaling of negativity in the Kondo regime.
This scaling may be regarded as yet another independent evidence
of the fact that $L^*$ is indeed the Kondo length $\xi$. In
general, the entanglement $E$ between the impurity and block $B$
is a function of the three independent variables, $J',L$ and $N$
which, due to the one to one correspondence between $J'$ and
$L^*$, can be written as $E(L^*,L,N)$. We find that, in the Kondo
regime, $E=E(N/L^*,L/N)$.  To illustrate this, we fix the ratio
$N/L^*$ and plot the entanglement in terms of $L/N$ for different
values of $J'$ (or equivalently $L^*$) for $J_2=0$ (Fig.
\ref{fig3}b) and for $J_2=J_2^c$ (Fig. \ref{fig3}c). The complete
coincidence of the two plots in Figs. \ref{fig3}b and c shows
that, in the Kondo regime, the spin chain can be scaled in size
without essentially affecting the entanglement as long as $L^*$ is
also scaled. In the dimer regime the entanglement stays a function
of three independent variables, i.e. $E=E(L^*,L,N)$, and, as shown
in Fig. \ref{fig3}d, the entanglement does not scale with $L^*$.
In our approach, the EHL $L^*$ may be evaluated in both the Kondo
and the dimer regime: the scaling behavior, as well as the
dependence of $L^*$ on $J'$, discriminates then between the very
different entanglement properties exhibited by the spin chain
Kondo model as $J_2$ crosses $J_2^c$.

We defined $L^*$ such that there is no entanglement between the
impurity and block $B$ when block $A$ is made of $L^*$ spins.
Conventional wisdom based on previous renormalization group
analysis suggests that, in both regimes, the impurity and the
block $A$ of length $L^*$ form a pure entangled state, while block
$B$ is also in a pure state. This is indeed approximately true in
the dimer regime (exactly true for $J_2=0.5$) but it turns out to
be dramatically different in the Kondo regime. To check this, we
computed the von Neumann entropy of the block $B$ when block $A$
has $L^*$ spins and found it to be non zero. Thus, the blocks $A$
and $B$ are necessarily entangled in the Kondo regime as there is
no entanglement between the impurity and $B$.
 In fact, after a distance
 $L^*$,
the impurity is "screened" i.e, the block $B$ feels as if it is
part of a conventional gapless chain and has a diverging von
Neumann entropy. The Kondo cloud is, then, maximally entangled
with the impurity as well as being significantly entangled with
block $B$. Based on the above, a simple ansatz for the ground
state $|GS\ra$ in the Kondo regime has been proposed in
\cite{Kondous}; namely, one may conjecture that
\begin{equation}\label{state_kondo}
    |GS_I\ra=\sum_i \alpha_i
    \frac{|\uparrow\ra|L_i^\uparrow(J')\ra-|\downarrow\ra|L^\downarrow_i(J')\ra}{\sqrt{2}}\otimes
    |R_i(J')\ra,
\end{equation}
where $\alpha_i$ are constants,
$\{|L_i^\uparrow(J')\ra,|L^\downarrow_i(J')\ra\}$ and
$\{|R_i(J')\ra\}$ are sets of orthogonal states on the cloud and
the remaining system, respectively. At the fixed point
$J'\rightarrow 0$ all spins except the impurity are included in
$|L_i^\uparrow(J')\ra$ and $|L_i^\downarrow(J')\ra$. At
$J'\rightarrow 1$, very few spins are contained in
$|L_i^\uparrow(J')\ra$ and $|L_i^\downarrow(J')\ra$ while
$\{|R_i(J')\ra\}$ represents most of the chain.

\begin{figure}
\centering
    \includegraphics[width=8cm,height=7cm,angle=0]{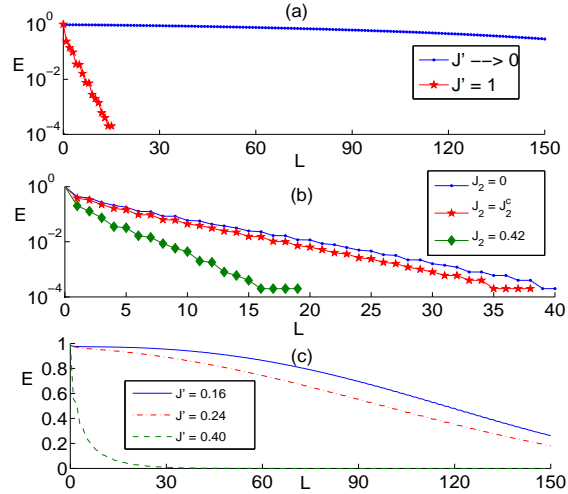}
    \caption{(a) Entanglement vs. $L$ for the two fixed points $J'\rightarrow 0$
     and $J'\rightarrow 1$ in the Kondo regime $J_2=0$.
    (b) Exponential decay of entanglement in terms of $L$ in a chain of length $N=250$ for $J'=0.6$.
    (c) Non-exponential decay for small $J'$ in the dimer regime $J_2=0.42$.}
     \label{fig4}
\end{figure}

  For what concerns the mere evaluation of the amount of entanglement as $J'$ is varied, we plot, in Fig.
\ref{fig4}a, the negativity as a function of $L$ near by the two
fixed points, i.e. $J'\rightarrow0$ and $J'\rightarrow1$,
accessible in the Kondo regime: as expected, near $J'\rightarrow0$
(i.e, for large values of the Kondo screening length), the
entanglement remains large for rather large values of $L$ while,
for $J'\rightarrow 1$ (i.e. for a very small cloud) it decreases
rapidly with $L$. Fig. \ref{fig4}a (semi-logarithmic) shows that,
also at the extreme limits $J'\rightarrow0$ and $J'\rightarrow1$,
the entanglement decays exponentially with $L$ since this a
characteristic mark of the entanglement in the Kondo regime. This
exponential decay of entanglement is absent in the dimer regime:
Fig. \ref{fig4}b shows that, in the gapped dimerised regime, only
for rather large $J'$, the entanglement decays exponentially with
$L$ while, for small $J'$, the entanglement between the impurity
and block $B$ decays {\em slower than an exponential} as a
function of $L$ exhibiting even a plateau at short distances. This
latter feature is evidenced in Fig. \ref{fig4}c, and is consistent
with the emergence, for small $J'$, of long range valence bonds
between the impurity and far spins as a consequence of the onset
of the dimerised ground state \cite{Sorensen-Affleck}. In fact,
when $J'$ is small, $J_2J'$ becomes much less than $J_2^c$ and the
impurity forms valence bonds with distant spins, while the other
spins, since for them $J_2>J_2^c$, form a valence bond with their
nearest neighbor to preserve the dimerised nature of the ground
state: as a result, the impurity shares less entanglement with
nearby spins and fulfills its capacity of entanglement forming
valence bonds with the more distant spins in the chain.

\section{LONG-RANGE DISTANCE-INDEPENDENT ENTANGLEMENT IN THE KONDO REGIME} \label{dynamics}

Entanglement between a single impurity spin and a group of spins-
such as the ones inside the Kondo cloud emerging in the Kondo
regime accessible to the spin chain Kondo model- cannot be used as
a resource for computational tasks since manipulation of many
particles is an extremely difficult task. It is much more
convenient, instead, to use entanglement between distant
individual particles since, due to their localization, one may
more easily resort to unitary gates and measurements to control
these individual spins. In \cite{Sodano} we proposed a procedure
to convert the {\em useless} entanglement between the impurity and
the cloud into the {\em useful} entanglement between the two
ending spins of the Kondo spin chain.

To engineer the long range entanglement between distant individual
spins we took \cite{Sodano} the finite Kondo spin  chain
(\ref{Hamiltonian}) in its ground state $|GS_I\ra$ and, then, {\em
pertinently} quenched the coupling at the opposite end of the
impurity allowing for the dynamics to develop entanglement. We
showed that, in the Kondo regime, the entanglement between the two
ending spins oscillates between high peaks with a periodicity
determined by $J'$, while the dynamics is very fast (thereby
decoherence hardly gets time to act) and is robust against thermal
fluctuations. At variance, in the gapped dimer regime, the
dynamics is much slower, qualitatively different and, in finite
chains, it still generates some entanglement due to the
unavoidable end to end effects, which are drastically tamed if one
"cuts off" the impurity from the chain during the dynamics. In the
Kondo regime, cutting off the impurity has minimal effect on the
final entanglement between the ending spins since, here, the
process is induced by the emergence of the Kondo cloud
\cite{Sodano}.

\subsection{Entanglement Oscillations Induced by Local Quench Dynamics in the Kondo Regime } \label{oscillations}
Initially, the system is assumed to be in the ground state
$|GS_I\ra$ of $H_I$. A minimal quench modifies only the couplings
of the $N$th spin by the amount $J'$ (same as $J'$ in Eq.
(\ref{Hamiltonian})) so that $H_I$ is changed to
\begin{eqnarray}\label{hamil_NNN_quenched}
    H_F&=&J'(J_1\sigma_1.\sigma_2+J_2\sigma_1.\sigma_3+J_1\sigma_{N-1}.\sigma_N+J_2\sigma_{N-2}.\sigma_N)\cr
    &+&J_1\sum_{i=2}^{N-2}\sigma_i.\sigma_{i+1}+J_2\sum_{i=2}^{N-3}\sigma_i.\sigma_{i+2}.
\end{eqnarray}
Since $|GS_I\ra$ is not an eigenstate of $H_F$ it evolves as
$|\psi(t)\ra=e^{-iH_Ft}|GS_I\ra.$ An entanglement $E(N,t,J')$
between the ending spins emerges as a result of the above
evolution.

To compute $E(N,t,J')$, in \cite{Sodano} we computed the reduced
density matrix $\rho_{1N}(t)=tr_{\hat{1N}}|\psi(t)\ra\la\psi(t)|$
of spins $1$ and $N$ by tracing out the remaining spins from the
state $|\psi(t)\ra$. Then, we evaluated $E(N,t,J')$ in terms of a
measure of entanglement valid for arbitrary mixed states of two
qubits called the concurrence \cite{wootters}. We showed that
entanglement took its maximum value $E_m$ at an optimal time
$t_{opt}$ and at optimal coupling $J'_{opt}$ such that
$E_m=E(N,t_{opt},J'_{opt})$. We evidenced in \cite{Sodano} that
$J'_{opt}$ is not a perturbation of $J_1$ and $J_2$. If, as
expected from scaling in the Kondo regime \cite{Affleck-scaling,
Kondous}, the dependence on $N$ and $t$ can be accounted for by a
redefinition of $J'$ (equivalently $\xi$), then $t_{opt}$ and
$J'_{opt}$ cannot be independent quantities.

Our numerical analysis showed that, in the Kondo regime,
$t_{opt}\propto N$ and that $J'_{opt}$ yields $\xi=N-2$; since
$\xi\propto e^{\alpha/\sqrt{J'}}$ one gets
$$t_{opt}\propto N\propto e^{\alpha/\sqrt{J'_{opt}}}$$.
For our choice of $J',J_1$ and $J_2$ the spin-chain dynamics is
not analytically solvable and one has to resort to numerical
simulations. Here, for $N>20$, we used the time-step targeting
method, based on the DMRG algorithm introduced in
\cite{whiteDMRG}. For $N<20$, instead, one may resort to exact
diagonalization.

\begin{figure}
\centering
    \includegraphics[width=8cm,height=5.6cm,angle=0]{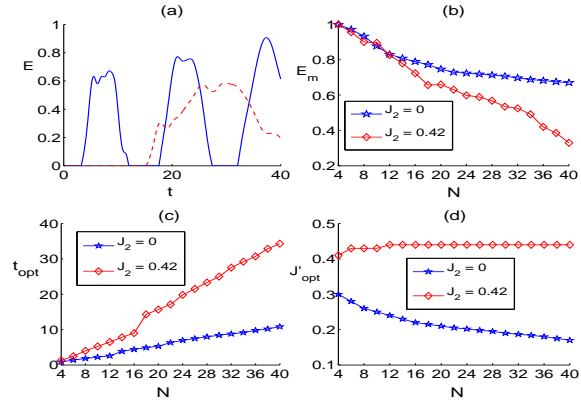}
    \caption{ Comparing the Kondo ($J_2=0$) and dimer ($J_2=0.42$) regimes.
    (a) Entanglement vs. time for $N=30$ with $J'=0.19$ for the Kondo regime (solid line) and
     $J'=0.44$ for the dimer regime (dashed line). (b) $E_m$ vs. the length $N$.
    (c) $t_{opt}$ vs. $N$. (d)  $J'_{opt}$ vs. length $N$.}
     \label{Fig5}
\end{figure}

We focused only on the first period of the entanglement evolution
in both regimes, since both decoherence and numerical errors make
it unwise to wait for longer times.
 Fig. \ref{Fig5}(a) shows that fast long-lived
(non-decaying) periodic oscillations with a period of $2t_{opt}$ characterize the time evolution in
the Kondo regime and that the maximal entanglement is achieved when the impurity coupling $J'$ equals
the value $J'_{opt}$ associated to a Kondo cloud of size $\xi=N-2$
(the Kondo cloud generated by the impurity sitting on the left side touches the other side of the chain);
in the dimer regime the dynamics appears
more dispersive and not oscillatory for any $J'$.
In Fig. \ref{Fig5}(b) we plot the maximum of entanglement, $E_m$, induced by bond quenching as a function of
the length $N$: though the entanglement decreases as $N$ increases, its value, in the Kondo regime,
stays rather high and becomes almost {\em distance independent} for very long chains; furthermore, as $N$ increases, the
entanglement generated in the Kondo regime is significantly bigger than the one
in the dimer regime.

Despite its lower value, achieving entanglement in the dimer
regime costs more time, as shown in Fig. \ref{Fig5}(c). It is also
clear from Fig. \ref{Fig5}(c) that $t_{opt}$ increases by $N$
linearly. At variance, in the Kondo regime, $J'_{opt}$ slowly
decreases as $N$ increases while it stays essentially constant in
the dimer regime (Fig. \ref{Fig5}(d)). This is commensurate with
the expectation that $t_{opt}\propto e^{\alpha/\sqrt{J'_{opt}}}$
in the Kondo regime, while, in the dimer regime, $t_{opt}$ and
$J'_{opt}$ are two independent quantities. The plot in Fig.
\ref{Fig6}a shows the exponential dependence of $t_{opt}$ on $J'^{
-1/2}_{opt}$ realized, in  the Kondo regime, for long enough
chains.

 \subsection{Probing the Size of the Kondo Cloud through Entanglement's Quench Dynamics  } \label{EntDyn}
 How the dynamics allows to engineer - even for very long chains of
size $N$- high entanglement oscillations between the ending spins
of the chain in the Kondo regime? To understand this, one should
recall that, in the Kondo regime, the impurity spin forms an
effective singlet with all the spins inside the cloud
\cite{Kondous} and that, only in this regime, one can always
choose $J'$ so that $\xi$ may be made comparable with $N$; at
variance, in the dimer regime, the impurity in $|GS_I\ra$ picks
out - no matter what the values of $J'$ and $N$ are- only an
individual spin in the chain to form a singlet ( a valence bond)
while the remaining spins form singlets ({\em local dimers}) with
their nearest neighbors \cite{Sorensen-Affleck}. Thus,
 only in the Kondo regime, one may use the remarkable resource to select - for any $N$- an initial state
 $|GS_I\ra$ which is free from local excitations: in fact, when $\xi=N-2$  ($J'=J'_{opt}$) there is only a {\emph single entity}, namely the $N$th spin,
 interacting with the impurity-cloud composite.
 As this situation can always be engineered by choosing, for any $N$, $J'=J'_{opt}$, this allows for an explanation \cite{Sodano} of the
distance-independent entanglement in Fig. \ref{Fig5}b. The
proposed scenario provides an intuitive grasp on why, only in the
Kondo regime, an optimal entanglement between the ending spins may
emerge from quench dynamics as the result of the interplay between
very few states. At variance, in the dimer regime, the energy
released by quenching is dispersed over the variety of different
quantum modes associated to the local dimers and no significant
long range entanglement may then be engineered.

To provide a more quantitative analysis, one may
expand $|\psi(t)\ra$ in terms of eigenvectors of $H_F$. By exact diagonalization (up to $N=20$), one finds that, in the
Kondo regime, only two eigenstates of $H_F$ (the ground state $|E_1\ra$ and one excited state $|E_2\ra$) predominantly contribute to the dynamics:
\begin{eqnarray}\label{Eigenvectors}
|E_1\ra&=&\alpha_1|\psi^-\ra_{1N}|\phi^-\ra_b+\beta_1(|00\ra_{1N}|\phi^{00}\ra_b\cr
&+&|11\ra_{1N}|\phi^{11}\ra_b
-|\psi^+\ra|\phi^{+}\ra_b)\cr
|E_2\ra&=&\alpha_2|\psi^-\ra_{1N}|\phi^-\ra_b-\beta_2(|00\ra_{1N}|\phi^{00}\ra_b\cr
&+&|11\ra_{1N}|\phi^{11}\ra_b
-|\psi^+\ra|\phi^{+}\ra_b).
\end{eqnarray}
In Eq. (\ref{Eigenvectors}) the first and the last spin are projected onto the singlet ($|\psi^-\ra$) and the
triplets ($|00\ra, |11\ra$ and $|\psi^+\ra$) while the states of all spins in the body of the chain have been
specified by the index $b$.
After a time $t$ the state evolves as
\begin{equation}\label{psit}
|\psi(t)\ra=\la E_1|GS_I\ra |E_1\ra+e^{-i\Delta Et}\la E_2|GS_I\ra |E_2\ra+...,
\end{equation}
where, $\Delta E$ is the energy separation between the two levels.
One defines $t=t_{opt}$ as the time for which the contribution of
$|\psi^-\ra_{1N}|\phi^-\ra_b$ is most enhanced in $|\psi(t)\ra$ due to
a {\em constructive interference}.

The condition for the onset of constructive interference is
\begin{equation}\label{interference}
|\la E_1|GS_I\ra\beta_1 | \approx |\la E_2|GS_I\ra\beta_2 |,
\end{equation}
so that terms other than $|\psi^-\ra_{1N}|\phi^-\ra_b$ in
$|\psi(t)\ra$ do not contribute at $t=t_{opt}$. When
$J'\rightarrow 1$ (very small cloud) then $|\la E_1|GS_I\ra|
\approx 1$ while $|\la E_2|GS_I\ra| \approx 0$ as the ground state
is hardly changed on quench: thus, constructive interference
between $|E_1\ra$ and $|E_2\ra$ it is impossible. The condition of
Eq. (\ref{interference}) cannot be satisfied also when
$J'\rightarrow 0$, since one has now that $\beta_1 \approx 0$ and
$\beta_2 \approx 1$ (the end spins form a singlet and triplet with
each other in $|E_1\ra$ and $|E_2\ra$ respectively
\cite{lorenzo1-illuminati}). Thus, only for intermediate $J'$
entanglement may peak. When $J'>J_{opt}$ ($\xi<N-2$)- and
particularly for $\xi<N/2$- the entanglement between the ending
spins is frustrated by the existence of local excitations whose
number increases as the size of the cloud gets smaller. In
addition, when $J'<J_{opt}$ ($\xi>N-2$), the Kondo cloud overtakes
the chain and the $N$th spin is already {\em included in the
cloud} and its tendency is to screen the original impurity as in
$|\psi^-\ra_{1N}|\phi^-\ra_b$ rather than to pair with it to form
a spin one (as in the last three terms of $|E_1\ra$). This makes
$\beta_1$ quite small, and it becomes smaller as the cloud
overtakes the chain and again the condition of Eq.
(\ref{interference}) cannot be fulfilled.
 Consequently, the optimal situation is realized when $J'=J'_{opt}$
 ($\xi=N-2$
 ), i.e just before the cloud overtakes the chain.
Thus, only in the Kondo regime, one can convert-for any $N$- the
\emph{useless entanglement} between the impurity spin and the
Kondo cloud into a \emph{usable entanglement} between the ending
spins of the chain. The emerging long distance entanglement
analyzed in this paper is, indeed, a genuine footprint of the
presence of the Kondo cloud in $|GS_I\ra$ (Fig. \ref{Fig7}(a)).

\begin{figure}
\centering
    \includegraphics[width=8cm,height=5.6cm,angle=0]{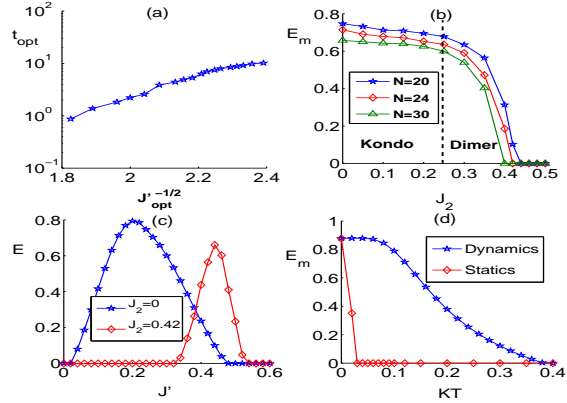}
    \caption{(a) $t_{opt}$ vs. $1/\sqrt{J'_{opt}}$ in the Kondo regime.  (b) $E_m$ at $t\propto 1/J'_{opt}$ vs. $J_2$ for chains of different lengths.
    (c) Entanglement attained at $t_{opt}$ vs. $J'$ for Kondo and dimer regimes for $N=20$.
    (d) $E_m$ vs. temperature after bond quenching (blue line)
    and induced by (static) weak coupling with the rest of the chain (red line) for a chain of $N=10$.}
     \label{Fig6}
\end{figure}

In Fig. \ref{Fig6}b we plot- for both regimes- the entanglement reached after waiting for a time interval of the
order of $1/J'_{opt}$. One notices that, for $J_2>J_2^c$, the entanglement peak decreases sensibly and goes
to zero rather soon. The plot of the maximal entanglement vs. $J'$ is given in Fig. \ref{Fig6}c: here one sees that, in
the Kondo regime, the entanglement rises from zero already at very small values of $J'$. This is expected since,
in the Kondo regime, to a small $J'$ is associated a large cloud containing the impurity sitting on the left side.

The essential role of the Kondo cloud in the entanglement
generation is further probed if one let evolve the ground state
$|GS_I\ra$ with a doubly quenched Hamiltonian obtained from
(\ref{hamil_NNN_quenched}) by isolating (i.e., putting $J'=0$) the
left hand side impurity while keeping
fixed to $J'_{opt}$ the bond connected to the last spin (see Fig.
\ref{Fig7}b). This forbids the dynamical build up of that
``portion" of the total entanglement which is only due to end
to end effects. In Fig. \ref{Fig8} we have plotted the  $E_m$ vs. $N$ after double quenching in
both regimes. Fig. \ref{Fig8} shows that entanglement in the dimer
phase collapses already when $N>12$ while
it stays \emph{unexpectedly} high - and almost independent on $N$ - in the Kondo regime; the existing entanglement
of the Kondo cloud with the impurity \cite{Kondous} is dynamically swapped over to the last spin.

 \subsection{Thermal Robustness of the Kondo Cloud Mediated Entanglement  } \label{thermal}

Note that a long distance singlet between the end spins may be
realized in a ground state when those spins are very weakly
coupled ($J'=\epsilon/\sqrt{N}<<1/\sqrt{N}$) to a spin chain
\cite{lorenzo1-illuminati}. This static approach to generate
entanglement relies on couplings which are so weak that they can
merely be regarded as perturbations. Such entanglement is not
robust against thermal fluctuations due to the smallness of the
gap ($\propto J'^2=\epsilon^2/N$) between the ground state and a
triplet state between the end spins. On the other hand our
approach enables to generate entanglement dynamically even for
$J'$ as high as $J'_{opt}\approx 1/(\log{N})^2$. Even when
temperature is increased, the entanglement is not seriously
disrupted till $K_BT$ exceeds the Kondo temperature ($\propto
1/\xi=1/(N-2)$) after which the Kondo cloud does not form. As a
result, while in the dynamical approach $K_BT< 1/(N-2)$, in the
static approach one has $K_BT<\epsilon^2/N$: thus, the long
distance entanglement generated through the dynamical approach is
thermally more stable. For instance, for $\epsilon\sim 10^{-1}$,
our dynamical approach is robust for temperatures 100 times higher
than those required for the static approach. In Fig. \ref{Fig6}d,
we plot $E_m$- as obtained in both approaches- vs. temperature for
$N=10$. In the static approach, the ground state is replaced by
the thermal state, whereas in our dynamic approach it is the
initial state which is taken to be the relevant thermal state. We
ignore thermalization and relaxation during dynamics since the
dynamical time scale, set by $t_{opt}$, is fast enough (this is
also an advantage over slow dynamical schemes with weak couplings
\cite{lorenzo2-plenio-li-wojcik}).

\begin{figure}
\centering
    \includegraphics[width=7cm,height=4.3cm,angle=0]{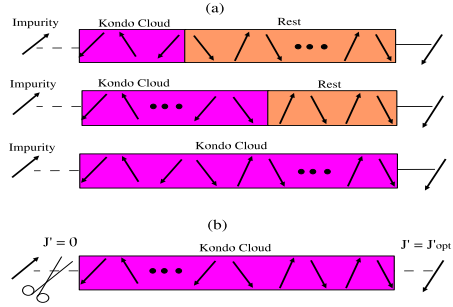}
    \caption{(a) Different $|GS_I\ra$ for entanglement generation through quench dynamics.
    Top: The ground states with $\xi<N/2$ (no entanglement). Middle: $N/2<\xi<N-2$ (some
    entanglement). Bottom:
    $\xi=N-2$ (optimal entanglement). (b) Decoupling the impurity from the chain. }
     \label{Fig7}
\end{figure}

\begin{figure}
\centering
    \includegraphics[width=6cm,height=4.5cm,angle=0]{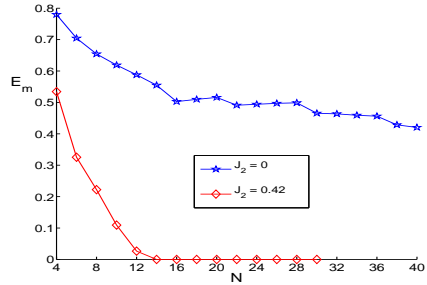}
    \caption{$E_m$ vs. $N$ after decoupling the first impurity.}
     \label{Fig8}
\end{figure}

We should also point out some systems where the Kondo
 cloud mediated long distance entanglement
 may be observed such as spin chains in ion traps \cite{Cirac-Porras-Wunderlich-Porras-Nature-2008},
 with trapped electrons \cite{Marzoli-Tombesi}, in chains of $P$ donors in $Si$ \cite{Kane} and Josephson
 chains with impurities \cite{glala}.

\section{Summary} \label{summary}

  To summarize, we reviewed our analysis \cite{Kondous} of the ground state entanglement of the Kondo spin
chain model from the viewpoint of a genuine entanglement measure,
namely the negativity. This readily showed that the impurity spin
is indeed maximally entangled with the Kondo cloud; we provided an
{\emph independent} method to determine the Kondo screening length
together with a characterization of the ground state of the Kondo
spin chain in the Kondo regime. We proposed an improved DMRG
approach enabling to account for the entanglement between the
impurity and a block of spins located at the other side of the
chain for different lengths of the block. We defined an
Entanglement Healing Length EHL and showed that, in the Kondo
regime, the EHL $L^*$ scales with the impurity coupling $J'$ just
as the Kondo length does. Finally, our approach showed that, in
the Kondo regime, the entanglement scales exponentially with
$L/L^*$ and that, in the gapped dimer regime, though it is still
possible to define an EHL, the impurity-block entanglement is
usually smaller and has no characteristic length scale.

We have also shown that substantial {\emph long range distance
independent} entanglement
 can be engineered by a non-perturbative quenching of a {\emph single} bond in the Kondo regime- and only here!- of a Kondo spin chain.
 This is the first example where a \textit{minimal local} action on a spin chain dynamically creates long range entanglement.
 In contrast to all known schemes for entanglement between individual spins in spin chains, here the entanglement attains a constant
 value for long chains rather than decaying with distance. This behavior arises since, in the Kondo regime, one may always select the
 initial state so as to fulfill the condition $\xi=N-2$.
 As the coupling is non-perturbative in strength ($J'_{opt}\approx 1/(\log{N})^2$), the entanglement generation
 is both fast and thermally robust. The
 long distance entanglement mediated by the cloud and the periodic
 dynamics, provides a clear
 signature of the existence of the Kondo cloud in a quantum system with
 impurities. These features are not seen when the Kondo cloud is absent, e.g., in the gapped dimer regime and- more remarkably- in
 other conventional gapless systems.

{\emph Acknowledgements--} AB and SB are supported by the EPSRC,
QIP IRC (GR/S82176/01), the Royal Society and the Wolfson
Foundation. PS was partly supported by the ESF Network INSTANS.

\end{document}